\documentclass[12pt]{article}

\usepackage{epsfig,amsfonts}

\def \be {\begin{equation}}
\def \ee {\end{equation}}                                         
\def \ba {\begin{eqnarray}}
\def \ea {\end{eqnarray}}

\begin{document}

\title{
On the SU(2)-Higgs Phase Transition
}
\author{Isabel Campos \thanks{Supported 
by a Fellowship from \emph{Ministerio de Educaci\'on y Cultura}, Spain}}
\bigskip
\maketitle

\begin{center}
{\it Departamento de F\'{\i}sica Te\'orica, Facultad de Ciencias,\\
Universidad de Zaragoza, 50009 Zaragoza, Spain \\
\small e-mail: \tt isabel@sol.unizar.es} \\
\end{center}
\bigskip
\begin{abstract}

The properties of the Confinement-Higgs phase transition in the 
SU(2)-Higgs model with fixed modulus are investigated. 
We show that the system exhibits a transient behavior up to {\it L}=24 
along which,  
the order of the phase transition cannot be discerned.
To get stronger conclusions about this point, without going to
prohibitive large lattice sizes, we have introduced a second 
(next-to-nearest neighbors) gauge-Higgs
coupling ($\kappa_2$). On this extended parameter
space we find a line of phase transitions which become
increasely weaker as $\kappa_2 \rightarrow 0$.
The results point to a first order character
for the transition with the standard action ($\kappa_2 = 0$). 

\end{abstract}

\newpage

\section{Introduction}

The generation of mass in the electroweak sector of the Standard Model (SM)
\cite{WEI} relies on the Higgs mechanism. Because of this fact large
amount of work has been spent, both, perturbative and non-perturbatively
in order to understand the continuum limit of gauge-Higgs models.
Underlying is the very question: Does a non-trivial QFT exists in the
limit of infinite cut-off ? (see \cite{CALLA} for a review).
 
On the one hand, it is almost rigorously proved \cite{AIZ,FRO} that the pure
scalar sector, namely the $\lambda \Phi^4$ model, leads to a trivial
theory when the cut-off is removed. Of course this is an academic model,
and the question is if the coupled gauge-scalar model could produce
a non-trivial theory in the continuum.

Perturbatively, the answer seems to be negative, only theories
asymptotically free in all the couplings can be constructed \cite{PHAS}.
But realistic theories have at least one coupling which is
not asymptotically free. A quite accepted scenario describes the
SM as an effective theory, with a finite cut-off
above which the theory is no longer valid. Within this approach
an upper-bound for the Higgs mass can be calculated (see \cite{AHAS}
for a review).

However, the non-perturbative sector of the SM is not yet
completely understood. The strong self-coupling allowed for
the Higgs field, renders perturbation theory useless, and one
is forced to use non-perturbative methods to get insight into the
properties of the model in this region of couplings.

The Higgs sector of the SM can be approximated by the SU(2)$\otimes$U(1)
Higgs model. This approximation is expected to behave reasonably well
for the Yukawa couplings between the fermions and the Higgs field,
excepting the coupling of the $top$ quark, are small. Also,
since the U(1) and the SU(2) gauge couplings are related through
the Weinberg angle 
by $g_{\mathrm{U}(1)} \approx 0.27 g_{\mathrm{SU}(2)}$, we can
start with the SU(2)-Higgs model, neglecting the U(1) degree of
freedom, as a first approximation to describe the electroweak
interaction.

In particular, we shall be interested in the limit in which the
modulus of the Higgs field is frozen. In the usual notation
it correspond to the limit $\lambda=\infty$, being $\lambda$ the
parameter controlling the radial degree of freedom of the Higgs field.
The phase diagram for this model is well known \cite{DROU,FRAD,CREU}. 
A phase transition (PT) line separates a region
where the scalar particles are confined in bounded states (confined phase),
from another region where the symmetry SU(2)$\otimes$SU(2) is
spontaneously broken, and the spectrum consists of the W's gauge bosons 
and the Higgs particle (Higgs phase).
In this sense SU(2)-Higgs is similar to QCD: there is a phase in which
gauge color fields form glueballs and quarks are confined into 
hadrons, and another phase characterized by the onset of the Higgs mechanism.

\begin{figure}[h]
\begin{center}
\epsfig{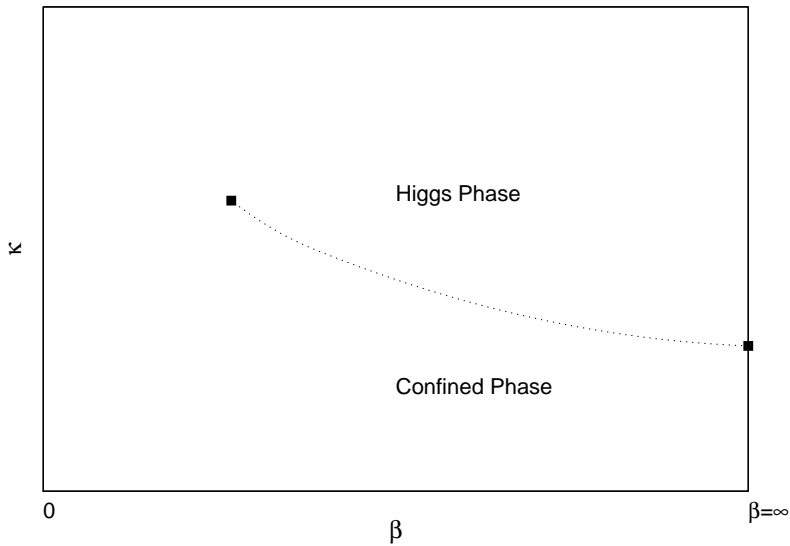}
\caption{\small {Schematic representation 
of the phase diagram of the SU(2)-Higgs 
model with frozen modulus ($\lambda = \infty$).}}
\label{SU2}
\end{center}
\end{figure}

The PT line ends at some finite point of the parameter space being
both regions analytically connected. So, strictly speaking, we should
talk about a single phase, the Confined-Higgs phase (see Figure \ref{SU2}).

In the scaling region, 
this model has been extensively studied for small \cite{LANG} and
intermediate \cite{LANG,MONT,BOCK} values of $\lambda$, where
the transition is distinctly first order. The PT weakens when 
increasing $\lambda$. In the limit $\lambda = \infty$
it is generally believed that the transition is still first
order, though extremely weak. However, 
in our opinion, a higher statistic study in the 
limit $\lambda = \infty$ is still lacking since the results up
to date are not conclusive, as we shall demonstrate. To shed some light
on this problem we have studied the model in an extended parameter
space too. However, our motivation is not only performing 
such a large statistics study,
but also we want to extract general properties of weak
first order phase transitions in coupled gauge-Higgs systems.
With this purpose, we have added
an extra positive gauge-Higgs coupling between next-to-nearest neighbors
to the standard action. We shall use it as a parameter
to study the weakening of the PT when this
extra coupling is tuned to zero.

In the next section we describe the model, and summarize previous
results. In section 3 the numerical method and analysis techniques
are detailed. Section 4 contains the results. Finally, the last
section is devoted to conclusions. 

\section{The Model}

The SU(2) lattice gauge model coupled to an scalar field, in the
fundamental representation of the gauge field can be described by the
action

\ba
S_{\lambda} & = \displaystyle \beta \sum_{p} [1 - 
\frac{1}{2} {\mathrm{Tr}}U_p] - \frac{1}{2} \kappa_1 \sum_{x, \mu} 
{\mathrm{Tr}}\Phi^{\dagger}(x) U_{\mu}(x) \Phi(x + \mu) + \nonumber \\ 
& \displaystyle  + \lambda \sum_{x} [\Phi^{\dagger}(x) \Phi(x) - 1]^2 + \sum_{x} \Phi^{\dagger}(x) \Phi(x)
\ea

Where U$_{\mu}(x)$ represents the link variables, and $U_p$ are their products
along all the positive oriented plaquettes of a four-dimensional
lattice of side {\it L}.

The scalar field at the site $x$ is denoted by $\Phi(x)$, being
$\lambda$ the parameter controlling its radial mode. 
In the limit ($\lambda = \infty$, $\beta = \infty$) the model becomes a pure
O(4)-symmetric scalar model.

As we pointed out in the previous section, the PT line ends at some
finite value of the parameters ($\beta$, $\kappa_1)$
The endpoint moves towards larger $\beta$ values as $\lambda$ increases.
For $\lambda \geq 0.1$ the endpoint crosses to the $\beta > 0$ region,
and in the limit $\lambda = \infty$ the phase transition 
ends at ($\kappa_1 \approx 0.6$, $\beta \approx 1.6$) \cite{LANG}.
It is commonly believed that the transition at this point is second order with
classical critical exponents \cite{ROB}, however a careful numerical
study would be necessary.

In the scaling region, and for finite $\lambda$, the phase transition 
turns out to be first order. Also,
the transition becomes weaker as $\beta$ or $\lambda$ increases. In
particular in the limit $\beta = \infty$ (spin model)
the transition is second order with classical critical exponents
\cite{ITZ}.

In the limit $\lambda = \infty$ the situation is less transparent,
and it is not clear whether the transition is weak first order or higher
order.

The study of the model with the parameter $\lambda = \infty$ is equivalent
to fixing the modulus of the Higgs field, $\Phi^{\dagger} \cdot \Phi = 1$. 
In the pioneer work, \cite{LRV}, the PT was considered first order for finite
$\lambda$ and second order for $\lambda = \infty$. Later,
larger statistics, and the hypothesis of a universal behavior of
the PT for all values of $\lambda$, seem to point to a first order 
character of the PT in the scaling region. Nowadays,
though it is generally believed that the transition is still first
order in this limit, the numerical proofs \cite{LM,MONT} 
on which rely these statements are not conclusive, as the
authors safely conclude, because the statistic and lattice sizes
are not enough for excluding the possibility of a higher 
order phase transition in the edge $\lambda = \infty$.

The model with fixed modulus is described by the action
\be
S_{\infty} = \beta \sum_{p} [1 - \frac{1}{2} {\mathrm Tr}U_p]
- \frac{1}{2}\kappa_1 \sum_{x, \mu} {\mathrm Tr} \Phi^{\dagger}(x) U_{\mu}(x) \Phi(x + \mu)  
\ee

As we shall show below, 
we have simulated this model up to lattices {\it L}=24, and the result
is still compatible with a second or higher order PT, however,
we have no indications of asymptoticity
in the behavior of the observables, and the results are compatible with
a very weak first order PT too. This means that larger lattices
are needed to overcome these transient effects,  
but the added difficulty here is that such lattices would suffer
of termalization problems, and severe autocorrelation times.
Altogether makes this approach too CPU expensive for nowadays computers.

In order to get a more conclusive answer without going to prohibitive
large lattice sizes, we have studied the model in an extended parameter space. 
For this purpose
we have introduced a second coupling between the gauge and the scalar
field connecting next-to-nearest neighbors on the lattice,
in such a way that the new action reads
\be
S = S_{\infty}
- \frac{1}{4} \kappa_2 \sum_{x, \mu<\nu}
Tr \Phi^{\dagger}(x) [U_{\mu}(x) U_{\nu}(x+\mu) +
U_{\nu}(x) U_{\mu}(x+\nu)] \Phi(x + \mu + \nu)     
\label{ACTION}
\ee

Within this parameter space we expect to get a global vision on what
the properties of the PT are, and also, to give a stronger conclusion 
about the order. In the region of
$\kappa_2$ positive, (competing interaction effects appear if $\kappa_2 < 0$) 
this extended model is expected to belong
to the same universality class than the {\it standard} one ($\kappa_2$ = 0) 
since both models posses the same symmetries.
The effect of the new coupling $\kappa_2 > 0$ is to reinforce the
transition but should not change the order if the system has not
tricritical points. An example of such behavior appears in the 
O(4)-symmetric $\sigma$ model with second neighbors coupling. This model 
presents a ($\kappa_1^c,\kappa_2^c$) line of phase
transitions which is second order, since the model with $\kappa_2=0$
shows a second order PT too \cite{YO}. 

The action (\ref{ACTION}) has the following symmetries:
\begin{itemize}
\item{} $\kappa_1 = \kappa_2 = 0$.
\ba
\beta  & \rightarrow & -\beta \nonumber \\
U_{\mu}(x) & \displaystyle \rightarrow & (-1)^{\sum_{\rho \neq \mu} x_{\rho}} U_{\mu}(x) 
\ea
\item{} $\kappa_1 \rightarrow -\kappa_1$, $\Phi(x)$ fixed
\be
U_{\mu}(x) \rightarrow - U_{\mu}(x) 
\ee
\item{} $\kappa_1 \rightarrow -\kappa_1$, $U_{\mu}(x)$ fixed
\be
\displaystyle \Phi(x) \rightarrow \displaystyle (-1)^{\sum_{\mu=0}^{d-1} x_{\mu}} \Phi(x) 
\label{STAG}
\ee
\end{itemize}

The action is not symmetric under the
change $\kappa_2 \rightarrow -\kappa_2$. The existence of couplings
$\kappa_1$ and $\kappa_2$  
with opposite signs would make frustration to appear, and
very different vacua are possible \cite{YO}. This region present
problems when one tries to implement reflection positivity, however, 
the possibility of defining a continuum limit at this region is not
discarded a priori \cite{POLO}. 
In this work we are interested in the regions free of frustration
effects. Taking into account the symmetry properties of the action,
the phase diagram in the region $\kappa_2 >$ 0 will be symmetric
with respect to the axis $\kappa_1$ = 0, and hence, we can restrict
the study to the quadrant $\kappa_1 >$ 0.

We define the normalized energy associated to the plaquette term
\be
E_0 = \frac{1}{N_{l_0}} \sum_{p} (1 - \frac{1}{2}{\mathrm{Tr}} U_p)
\ee
and also the energies associated to the links
\be
E_1  = \frac{1}{N_{l_1}} \sum_{x,\mu} \frac{1}{2} {\mathrm{Tr}} \Phi^{\dagger}(x) U_{\mu}(x) \Phi(x + \mu)  
\ee
\be
E_2  =  \frac{1}{N_{l_2}} \sum_{x, \mu<\nu} \frac{1}{4}
{\mathrm{Tr}} \Phi^{\dagger}(x) [U_{\mu}(x) U_{\nu}(x+\mu) +
U_{\nu}(x) U_{\mu}(x+\nu)] \Phi(x + \mu + \nu)   
\ee

where $N_{l_0}$ = 6V, $N_{l_2}$ = 12V and $N_{l_1}$ = 4V.

With these definitions $E_0 \rightarrow 0$ when $\beta \rightarrow \infty$
and $E_i \rightarrow 1$ when $\kappa_i \rightarrow \infty$.

On the three-dimensional ($\beta$, $\kappa_1$, $\kappa_2$)
parameter space we consider the plane $\beta=2.3$. 
On this plane there is a PT line ($\kappa_1^c, \kappa_2^c$).
We expect to learn on the properties of this PT on the region
($\kappa_1^c \neq 0, \kappa_2^c \neq 0$), where the signals are
clearer, with regards to applying what we learn to the $\emph standard$
case $\kappa_2=0$. 

To monitorize the strength of the phase transition, we measure
the existence of latent heat, and the behavior of the specific heat.
As we shall see, 
for $\kappa_1 = 0$ the transition is first order, with a
clearly measurable latent heat. We will see how this transition weakens
along the PT line for increasing values of $\kappa_1$. 

\begin{figure}[t]
\begin{center}
\epsfig{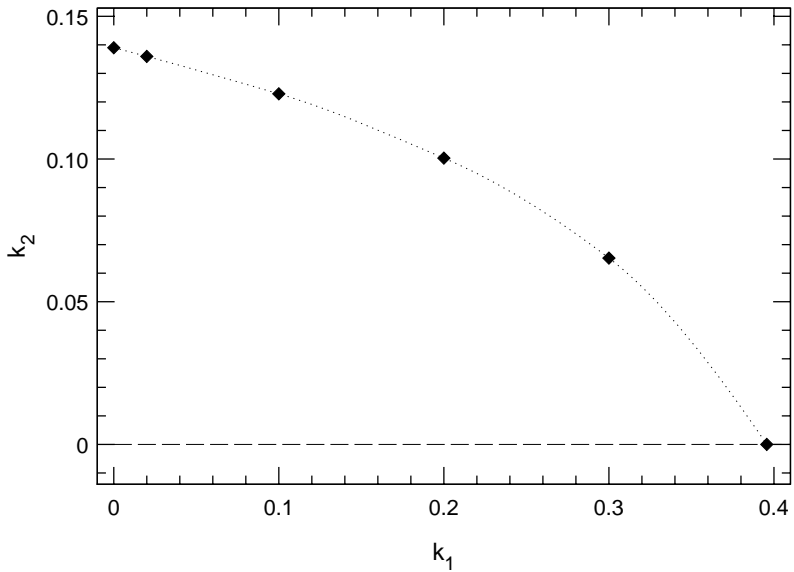}
\caption{\small {Phase diagram obtained from the MC simulation.}}
\label{PHASE}
\end{center}
\end{figure}           

\section{Numerical study}

\begin{figure}[t]
\begin{center}
\epsfig{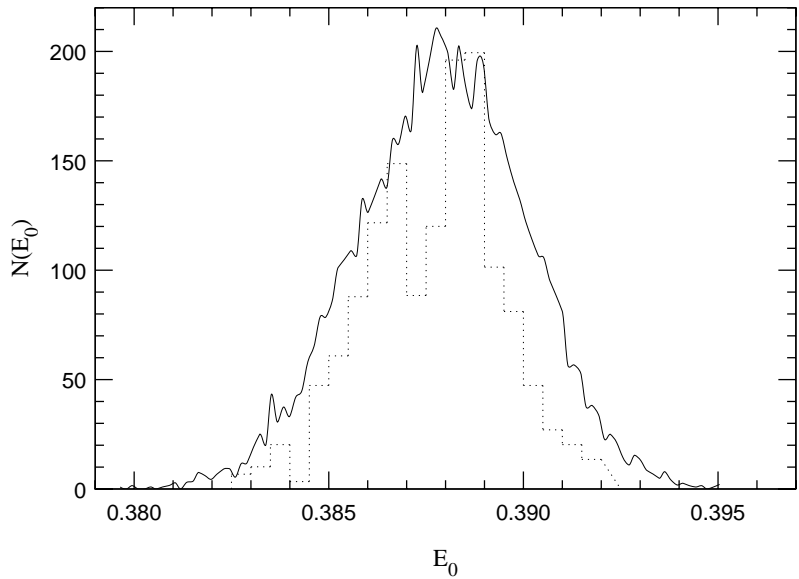}
\caption{\small {Normalized energy distribution for L=12, 
$\lambda=\infty$, $\kappa_1$=0.395, $\beta$=2.3 from \protect\cite{LM} 
(dotted line), compared with the distribution we obtain at the same couplings 
(solid line) in $L$=12 too, when statistics is increased by 
one order of magnitude.}}
\end{center}
\label{LMf}
\end{figure}

We have simulated the model in a $L^4$ lattice with periodic boundary
conditions. For the update we have employed a combination of
heat-bath and over-relaxation algorithms (ten over-relax sweeps
followed by a heat-bath sweep). For the simulation we used the RTNN
machine, consisting of a network of 32 PentiumPro 200MHz. processor.
The total CPU time employed has been the equivalent of 3 years of PentiumPro.

Monte Carlo methods provide information about the thermodynamic
quantities at a particular value of the couplings. We have used
the Spectral Density Method (SDM) \cite{SDM} to extract information
on the values of the observables in a finite region around the
simulation point. In particular it is useful to have a precise
location of the coupling where some observables have a maximum,
as well as an accurate measure of the value of that maximum.

From the Monte Carlo simulation at some coupling $\kappa$, 
we got the histogram $H(E)$ which is an 
approximation to the density of states. Using the SDM approximation
the probability of finding the system with
an energy E at a different coupling $\kappa^{\prime}$ can
be written as:
\be
P_{\kappa^{\prime}}(E) \propto 
H(E) e^{(\kappa^{\prime} - \kappa) V E}
\label{SD}
\ee

The region of validity of the SDM approximation
is $\Delta \kappa \sim 1/(V\sigma)$, being $\sigma$ the width
of the distribution $H(E)$. 
Although $\sigma$ gets the maximum values in the critical region, 
the approximation has been very useful, specially for tuning the
couplings where to measure.

Concerning the lattice sizes, we have used lattices ranging 
from $L$=6 to $L$=24. For the small lattices ($L$=6, 8 and 12) we have 
done 4$\times$10$^5\tau$ iterations, being
$\tau$ the largest autocorrelation time for the energy, 
which ranges from $\tau \approx 10$
in $L$=8 to $\tau \approx 35$ in $L$=24.
For the largest lattices, $L$=20 and 24, we run up to 10$^5\tau$
MC iterations.

The statistical errors are computed with the jackknife method.

\section{Results}

We shall make the discussion with the first-neighbors link
energy, E$_1$, but as far as the critical behavior is concerned, 
we could carry out the analysis with any of the energies. We remark
that an appropriate linear combination of E$_1$, E$_2$ and E$_0$
could give slightly more accurate results \cite{U1H}.

We have considered fixed values of $\kappa_1$ (0, 0.02,
0.1, 0.2 and 0.3) and sought the $\kappa_2$ critical for
every line, $\kappa_2^c (\kappa_1)$.
We have also studied the case $\kappa_2$= 0 varying $\kappa_1$
which corresponds to the usual SU(2)-Higgs model.

The SDM has been used to locate the apparent critical point,
defined through the specific heat behavior. From the specific
heat matrix
\be
C_v^{i,j}(L) = \frac{\partial E_i}{\partial \kappa_j}
\label{MAT}
\ee
we obtain the best signal for C$_v^{1,2}(L)$, which can
be calculated as (we shall omit the superscript from now on)
\be
C_v(L) = 4 L^d (\langle E_1 E_2 \rangle - \langle E_1 \rangle \langle E_2 \rangle)
\ee

In a first order phase transition, C$_v^\mathrm{max}(L)$ behaves, 
asymptotically, proportional to the volume,  $L^d$.
If the PT is second order, the dominant behavior for
C$_v(L)^\mathrm{max}$ 
is $L^{\alpha /\nu}$ which diverges too provided that $\alpha >$ 0.
At the upper critical dimension $\alpha$ = 0,  and one has to go further
the leading order, appearing logarithmic divergences \cite{LOG}.

As a consequence of this divergent behavior, in a finite lattice C$_v(L)$ 
shows a peak at some value of the
coupling which will be taken as apparent critical point,  
$\kappa_2^{\ast}(L)$.

In Figure \ref{PHASE} we plot the critical line ($\kappa_1^c, \kappa_2^c$).
This line is obtained by extrapolation to the thermodynamic 
limit according to $\kappa_2^c (\infty) = \kappa_2^{\ast} (L) - A L^{-d}$.
We point out that this extrapolation is valid only in first order
phase transitions. If the PT is second order the power in $L$
is ($-1/\nu$).

In continuous PT scale invariance holds, and the
thermodynamic magnitudes, such as the specific heat, or susceptibilities
do scale.
However if the transition is first order the correlation
length remains finite and hence there is not scaling properties,
and no critical exponents can be defined.

Nevertheless in a first order PT we can ask how large is the lattice
we need in order to observe the asymptotic behavior of C$_v$. In particular, 
with abuse of language one can measure a $\emph pseudo$ $\alpha/\nu$
exponent to get insight on the nature of the PT: the larger is the lattice
we need to measure $\nu = 1/d$, the weaker the PT is. Following this,
first order PT can be classified according to their degree of weakness.

\begin{figure}[t]
\begin{center}
\epsfig{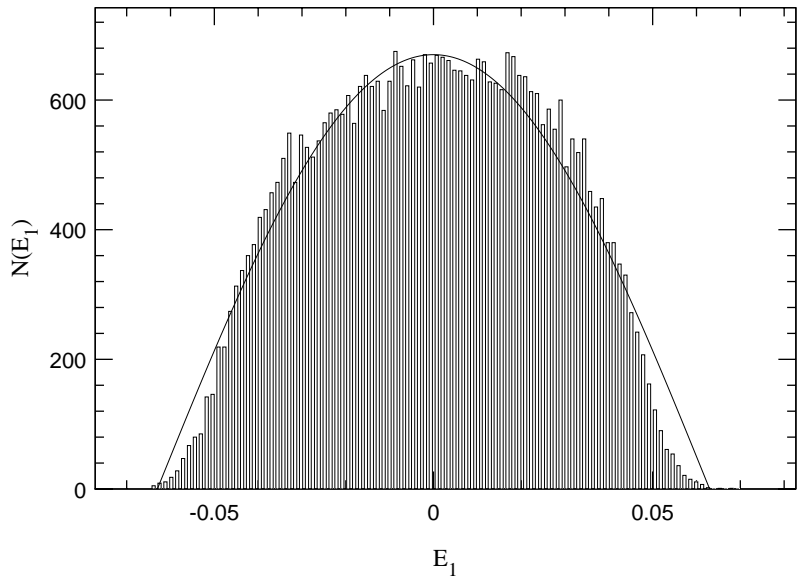}
\caption{\small 
{E$_1$ distribution on the axis $\kappa_1 = 0$ on a L=8 lattice.
at ($\kappa_2 = 0.15, \beta = 2.3$). The cosine fit is very accurate
in spite of the finite $\beta$ value, in this region of parameters
the pure gauge term couples slightly both sub-lattices.} }
\end{center}
\label{EJE}
\end{figure}

\begin{figure}[t]
\begin{center}
\epsfig{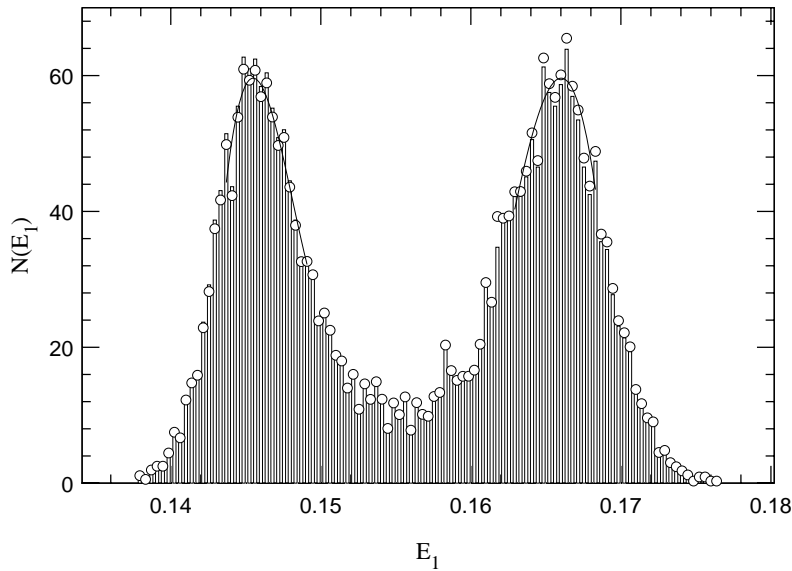}
\caption{\small {Normalized distribution of E$_1$ 
at ($\kappa_1=0.2$, $\kappa_2=0.10036$) in
a L=16 lattice. The cubic fit at the maxima to get
$\Delta E$ is superimposed.}}
\end{center} 
\label{FIT}
\end{figure}
      
\begin{figure}[t]
\begin{center}
\epsfig{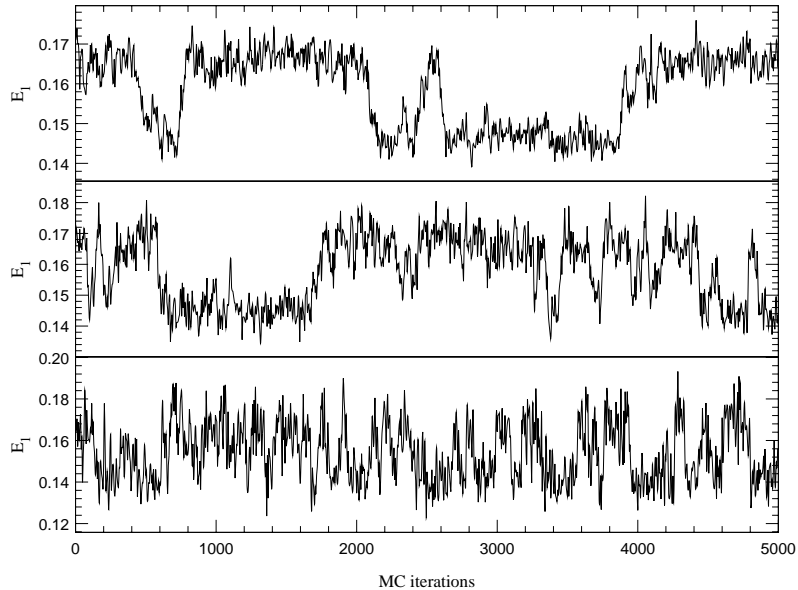}
\caption{\small {MC evolution of E$_1$ at ($\kappa_1=0.2, \kappa_2^{\ast}(L)$)
for $L$=8 (lower part), $L$=12 (middle) and $L$=16 (upper part).} }
\label{MET}
\end{center}
\end{figure}

\begin{figure}[t]
\begin{center}
\epsfig{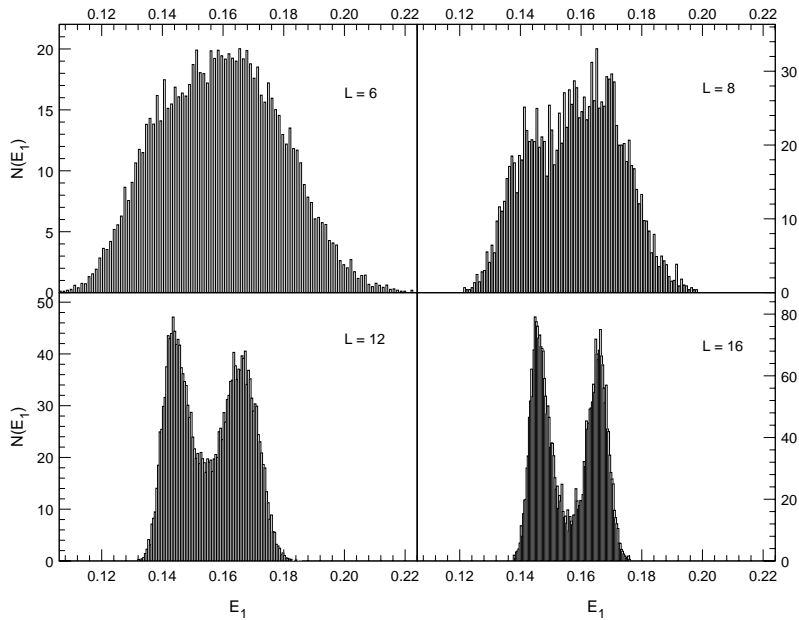}
\caption{\small {Normalized distribution of E$_1$ at 
($\kappa_1=0.2, \kappa_2^{\ast}(L)$)
for L=6, 8, 12 and 16.}} 
\label{HK02}
\end{center}
\end{figure}

However, the so called
weak first-order PT appear often in literature (see \cite{JUAN,LAF}
and references there in) as
PT characterized by a transient behavior with a non-measurable latent
heat. Let be $\xi_c$  the correlation length of the system
at the critical point in the thermodynamic limit. In a finite
lattice of size $L$, the first order behavior will be evidenced
if $L \geq \xi_c$. For lattice sizes much smaller than $\xi_c$ the system
will behave like in a second-order PT, since the correlation
length is effectively infinite. 
As an example, in Figure \ref{LMf} we plot the energy distribution
in a L=12 lattice at ($\beta=2.3, \kappa_1=0.395$) obtained in \cite{LM},
compared with the one we obtain at the same couplings and in the same L,
when the statistics increases by one order of magnitude. 
We observe that the order of the PT can not be discerned at this volume, 
even when the statistics is enough. Termalization effects can also
contribute to mistake the histogram structure.

The entire line ($\kappa_1^c, \kappa_2^c$) is first order, but the weak
character increases as $\kappa_2 \rightarrow 0$. We will make a quantitative
description of the weakening phenomenon
by studying the specific heat, and the latent heat.

But before going on, we shall make a remark concerning the behavior
on $\kappa_1 = 0$.
As we pointed out, the system is symmetric under the change $\kappa_1
\rightarrow -\kappa_1$. The transformation (\ref{STAG}) maps
the positive $\kappa_1$ semi-plane with energy E$_1$, onto the
negative $\kappa_1$ semi-plane with energy -E$_1$. The transition
across this axis is first order because the energy is discontinuous. 
In the limit $\beta \rightarrow \infty$, and in $\kappa_1 = 0$, the system
decouples in two independent sublattices, each one constituted by the first 
neighbors of the other. The first neighbors energy for this system 
is proportional
to $\cos \theta$, being $\theta$ the angle between the symmetry breaking
direction of the scalar field in both sub-lattices. In Figure \ref{EJE} we plot
the E$_1$ distribution for L=8 at $(\kappa_1 =0,\kappa_2 = 0.15)$ and
$\beta = 2.3$. We see that the agreement with a cosine distribution
is quite good in spite of the finite $\beta$ value.

\begin{figure}[t]
\begin{center}
\epsfig{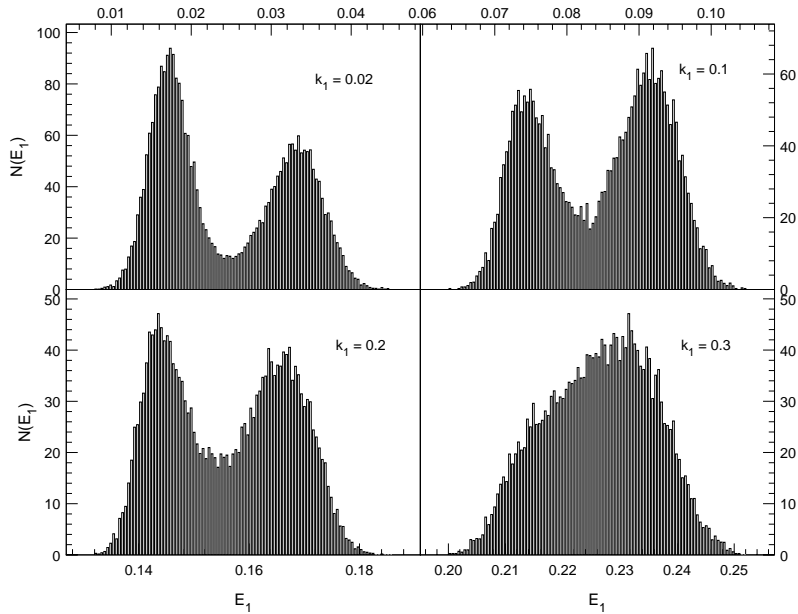}
\caption{\small {Normalized distributions of E$_1$ for 
$\kappa_1$ = 0.02, 0.1, 0.2 and 0.3 
at $\kappa_2^{\ast}(12)$.}} 
\label{H12}
\end{center}
\end{figure}

\begin{figure}[t]
\begin{center}
\epsfig{figure=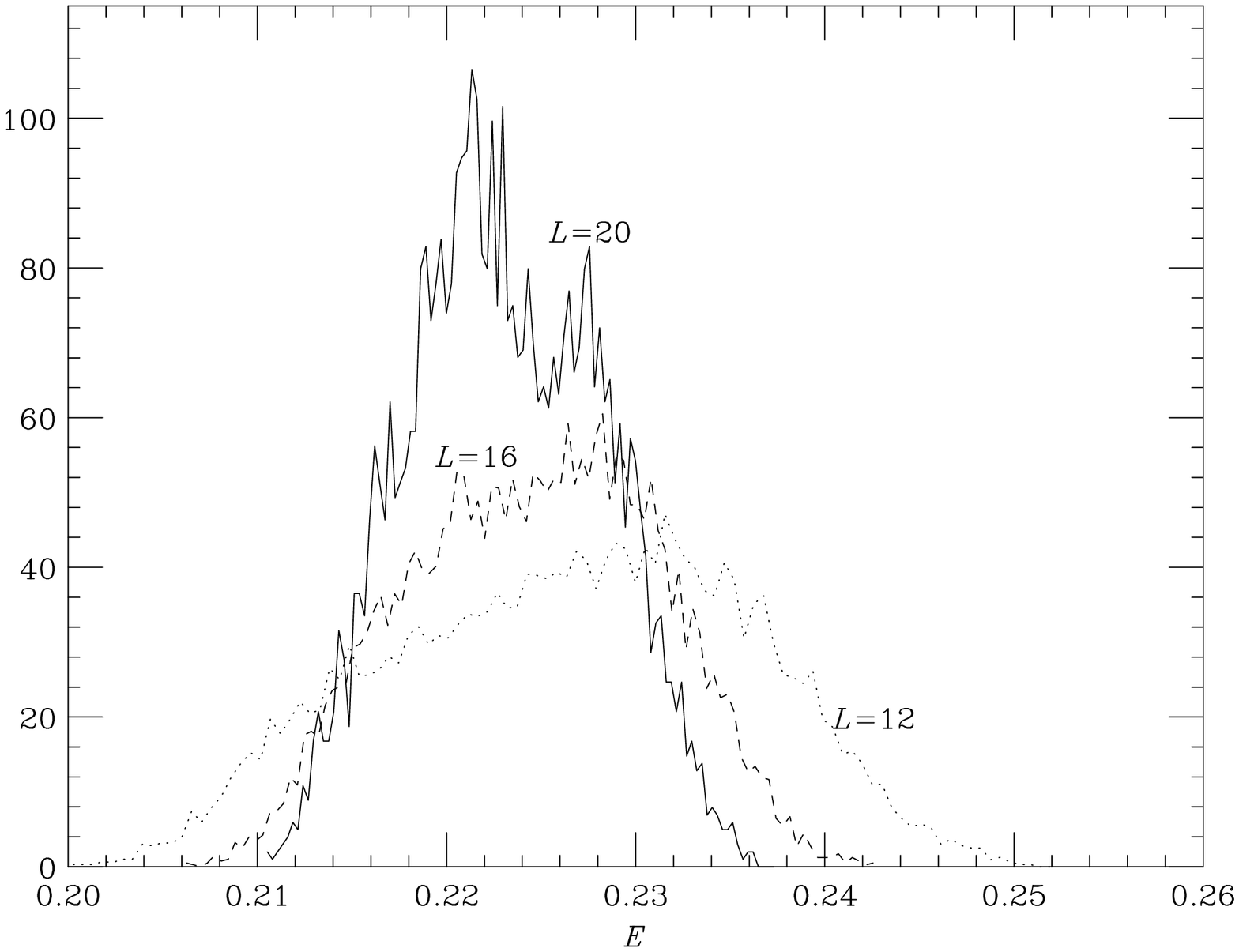,angle=90,width=300pt}
\caption{\small {Normalized distribution of 
E$_1$ at $\kappa_1$ = 0.3 for L=12,16 and 20}} 
\label{1620}
\end{center}
\end{figure}
\subsection{Latent Heat}

Along the apparent critical line we have done simulations for different
lattice sizes and stored the plaquette and links energies to construct the 
histograms for the energy distributions.
In a first-order phase transition the energy has a discontinuity 
which manifest in the appearance of latent heat, $\Delta E$. 
This quantity is not well
defined in a finite lattice, so we measure the distance between the
two maximum of the energy distribution, and extrapolate to
the thermodynamic limit. The drawback of this approximation is that
the maxima of the energy distribution are difficult to discern, since
this function at the apparent critical point is very noisy 
We have used a cubic spline at the maxima in order to get a 
more reliable estimation (Figure 5).

In Figure \ref{MET} we show the MC evolution
of E$_1$ for $L$=8, 12 and 16 at $\kappa_1 = 0.2$. 
In $L$=8 the latent heat is not clearly measurable. We observe in
the MC evolution how the two-state signal becomes cleaner as
the lattice size increases (see Figure \ref{HK02}).

In Figure \ref{HK02} we plot the distribution of E$_1$  
at $\kappa_1 = 0.2$ at the apparent
critical point $\kappa_2^{\ast} (L)$ for $L$=6, 8, 12 and 16.
A remarkable stability of $\Delta E_1$ with the volume is observed. This
is a common feature  for all values of $\kappa_1$.

As we have already pointed out, the transition weakens when increasing
$\kappa_1$ and larger lattices are needed in order to observe a
measurable latent heat. We give a quantitative
description of this fact in Figure \ref{H12}, where
the distribution of $E_1$ for several values 
of ($\kappa_1,\kappa_2^{\ast}(12)$) is displayed. 
The two-state signal is no longer measurable in $L$=12 at $\kappa_1 = 0.3$. 
The first evidences of two-states appear in $L$=20 (see Figure \ref{1620}).
but from its energy distribution we can only give an approximate value
for $\Delta E_1(L=20)$ since the two-peaks appear too close to each other.

\begin{figure}[t]
\begin{center}
\epsfig{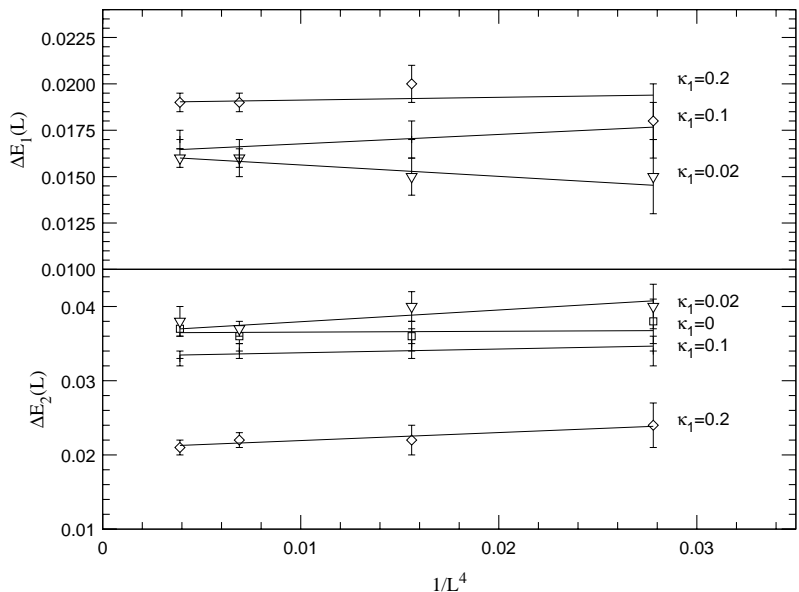}
\caption{\small {$\Delta E_1$ (upper plane)
and $\Delta E_2$ (lower plane) as a function of 1/L$^4$. The two-peak
structure is not clearly observed in L=6 at any $\kappa_1$ value.
The values quoted for this lattice size are upper bounds.}} 
\label{DELTA}
\end{center}
\end{figure}

In Figure \ref{DELTA} we plot $\Delta E_1(L)$ and $\Delta E_2(L)$  
as a function of 1/L$^4$, in order to get $\Delta E_i$ in the
thermodynamic limit with a linear fit.
Finally we quote these values in Table \ref{TAB}, together 
with the change in the action (\ref{ACTION}) between the two
phases.

\begin{figure}[h]
\begin{center}
\epsfig{figure=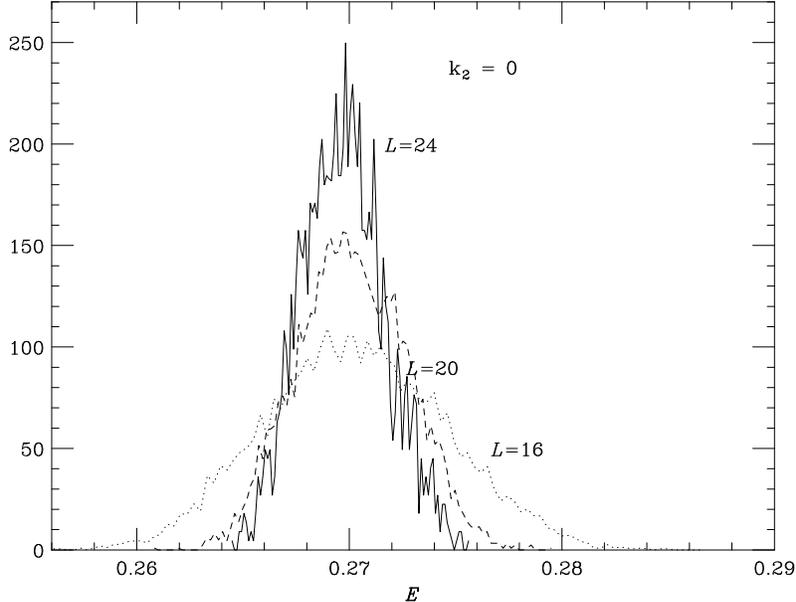,angle=90,width=300pt}
\caption{\small {Normalized distribution of E$_1$ at $\kappa_2$=0,
in $L$=16, 20 and 24 at the peak of the specific heat} }
\label{2024}
\end{center}
\end{figure}

\begin{table}[t]
{

\begin{center}
{
\begin{tabular}{|c|c|c|c|}\hline
coupling  &$\Delta E_1 (\infty)$ &$\Delta E_2 (\infty)$  &$\Delta S$ \\ \hline
$\kappa_1$ = 0             & -          &0.0366(8)  &0.0134(12) \\ \hline
$\kappa_1$ = 0.02          &0.0162(6)   &0.0347(5)  &0.0137(13) \\ \hline
$\kappa_1$ = 0.1           &0.0162(7)   &0.0345(9)  &0.0094(10) \\ \hline
$\kappa_1$ = 0.2           &0.0179(7)   &0.0201(8)  &0.0078(12)  \\ \hline
$\kappa_1$ = 0.3 & $\approx$0.006 &$\approx$ 0.012 &$\approx$ 0.0026 \\ \hline   
\end{tabular}
}
\end{center}
}
\caption[a]{\small {$\Delta E(\infty)$ for E$_1$ and E$_2$, and variation
of the action.}}
\protect\label{TAB}

\end{table}

From the energy distributions at $\kappa_2 = 0$, see Figure \ref{2024}
we have no direct
evidences of the existence of latent heat. However, on the larger
lattices one can observe non-gaussianities in the energy distributions. 
Such asymmetries  could precede 
the onset of clear two-peak structures in larger lattices, however
this is just a guess. We conclude that no information concerning the
order of the PT can be obtained from the energy distributions
up to $L$=24.

\subsection{Specific Heat}

We have done MC simulations at the points predicted by SDM, 
($\kappa_1, \kappa_2^{\ast}(L)$), in order to measure accurately
the peak of $C_v(L)$.

\begin{figure}[t]
\begin{center}
\epsfig{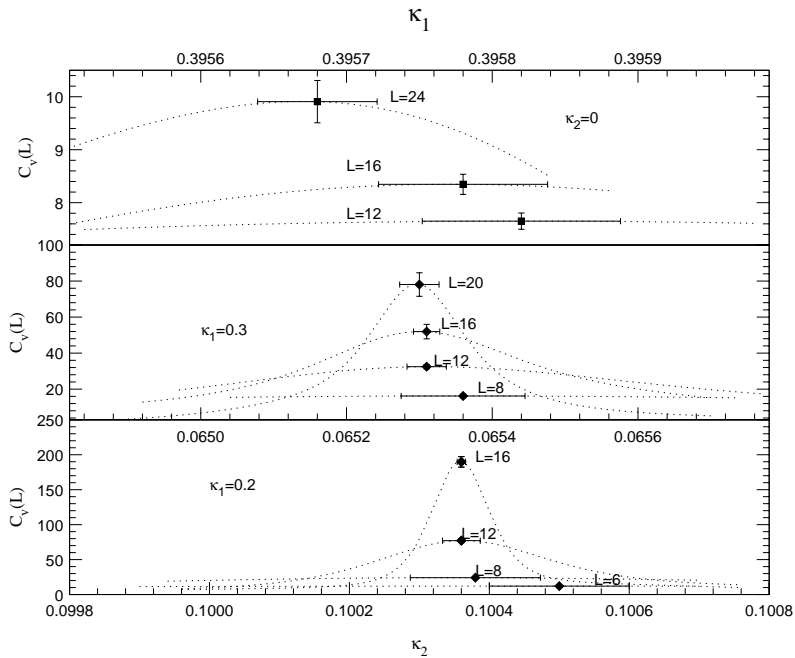}
\caption{\small {C$_v^\mathrm{max}(L)$ at $\kappa_1=0.2$ (lower plane),  at 
$\kappa_1=0.3$ (middle) and at $\kappa_2=0$ (upper plane).
The dotted line is the SDM extrapolation. } }
\label{PEAKS}
\end{center}
\end{figure}

As an example we show in Figure \ref{PEAKS} the value of $C_v(L)$ 
around its maximum for various lattice sizes at
$\kappa_1$ = 0.2 (upper plane) and at $\kappa_1$ = 0.3 (lower plane).
We observe that the maximum of the specific heat grows slower when
increasing $\kappa_1$, indicating a weakening in the PT. 

\begin{figure}[t]
\begin{center}
\epsfig{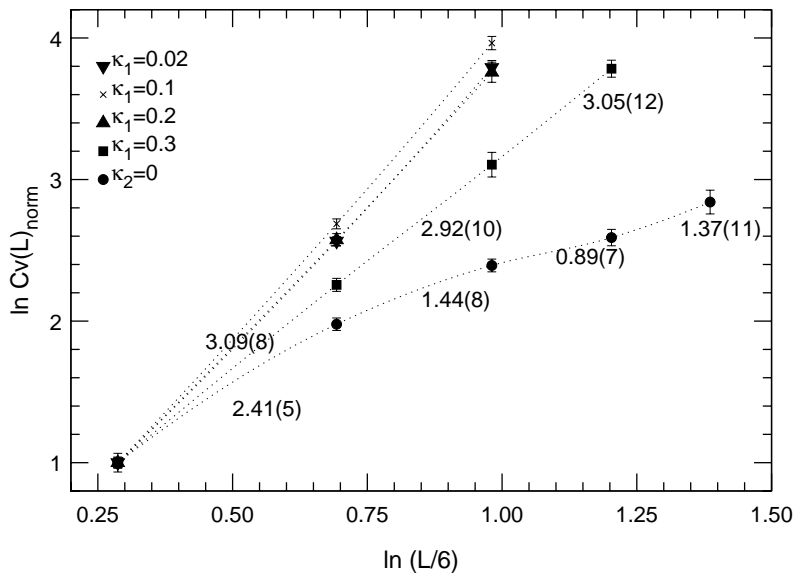}
\caption{\small {C$_v^\mathrm{max}$ for the 
various $\kappa_1$ values and $\kappa_2=0$.
We have normalized the values with respect to 
C$_v^\mathrm{max}$(8)/C$_v^\mathrm{max}$(6). 
The slope of the segments is indicated when smaller than 4.}} 
\label{CV1N}
\end{center}
\end{figure}

In Figure \ref{CV1N} we draw C$_v^\mathrm{max}(L)$ relative to 
C$_v^\mathrm{max}(6)$ as a function of the lattice size. 
The values have been normalized to 
C$_v^\mathrm{max}(8)$/C$_v^\mathrm{max}(6)$
in order to compare distinctly the behaviors for different $\kappa_1$.
The slope of the segment joining
the values of C$_v^\mathrm{max}$ in consecutive
lattices gives the $pseudo$ $\alpha/\nu$ exponent.
We observe that such slope is approximately 4 at $\kappa_1$ = 0.02, 
0.1 and 0.2 for all the volumes we compare.  However the transition
at $\kappa_1$ = 0.3 evidences much more weakness. We do not
have evidences of asymptoticity in C$_v^\mathrm{max}$ till L=20, as could
be expected from the energy distributions (Figure \ref{1620}).
The slope of the segment joining C$_v^\mathrm{max}(16)$ with 
C$_v^\mathrm{max}(20)$
is 3.05(12) which is almost the asymptotic value expected for
a first order PT.

As expected, only if the two-peak structure is
observed in the energy distributions and $\Delta E$ is stable,
the maximum of the specific heat will grow up like the volume, $L^d$.

At $\kappa_2$ = 0 we are within the transient region even for L=24. 
We remark that in this case C$_v$ is defined by the element C$^{1,1}$
of the specific heat matrix (\ref{MAT}) 
since this is the most natural choice
at this point, and also is the best signal we measure.
As we observe in Figure \ref{CV1N}, at $\kappa_2$ = 0 C$_v^\mathrm{max}(L)$
seems to tend to a constant value as V$\rightarrow \infty$ up to
$L$=20. From our previous discussions we should conclude that
either the correlation length at the transition point $\xi_c$
is much larger than the lattice size up to L=20, or the transition
is second order with $\alpha = 0$ in the thermodynamic limit. 
However, in $L$ = 24 things are changing, C$_v^{\mathrm max}(L=24)$
starts to run away of this quasi-plateau, and the $pseudo$ 
$\alpha/\nu$ exponent grows again.
As can be observed in Figure \ref{CV1N}, at $\kappa_1$ = 0.3 the $pseudo$ 
$\alpha/\nu$ exponent decreases for the segment $L$ = 12-16,
with respect to the value in the segment $L$ = 8-12. The lattice 
$L$ = 20 is enough to overcome the transient
region, but the behavior is qualitatively the same that in $\kappa_2$ = 0,
though the transition is stronger.

We believe that this behavior is general for weak first
order transitions in four dimensions. There exists a transient region in
which the correlation length is effectively infinite 
compared with the lattice size, and
the system behaves like suffering a second order PT with
thermal index $\alpha \approx 0$ in the thermodynamic limit.

\subsection{Binder Cumulant}

In order to check the consistency of our results 
we have also considered the behavior of the Binder cumulant

\be
V_L = 1 - \frac{\langle E_1^4 \rangle_L}{3\langle E_1^2 \rangle_L^2}
\label{CUMUL}
\ee

This quantity behaves differently depending on the order
of the PT. If the transition is second order the minimum
of the cumulant,
$V_L^{min}$ approach 2/3 in the thermodynamic limit.
However if the transition is first order, $V_L^{min}$ tends a value smaller
than 2/3 reflecting the non-gaussianity of the energy
distribution at the transition point.

\begin{figure}[t]
\begin{center}
\epsfig{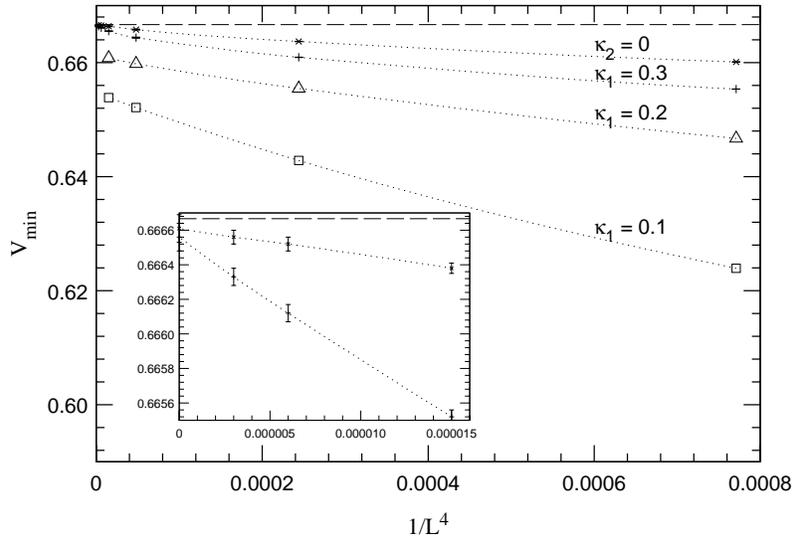}
\caption{\small {$V_L^{min}$ for $\kappa_1$ = 0.1, 0.2,0.3 and $\kappa_2$ = 0,
as a function of 1/L$^4$. The dashed line represents the value 2/3.}} 
\label{CUM}
\end{center}
\end{figure}

In Figure \ref{CUM} we plot $V_L^\mathrm{min}$ for several
$\kappa_1$ values.

For those values of $\kappa_1$ in which the PT is distinctly
first order, the minimum of the Binder cumulant stays
safely away from 2/3, as
we observe $V_L^{min}$ extrapolated to L$\rightarrow \infty$ is
0.65401(4) at $\kappa_1$ = 0.1. However, this value reachs
0.66637(5) at $\kappa_1$ = 0.3, and 0.66657(8) at $\kappa_2=0$.
Again we find a tight difference between a very weak first order PT
and a continuous one.

For the sake of discussing quantitatively the 
order of magnitude of the latent heat in the limit $\kappa_2=0$,
from the energy distributions we find that at $\kappa_2=0$
one can approximately locate one the peaks of the energy 
at E$_a \approx$ 0.273. The other should be at certain 
E$_b$ = E$_a - \Delta$, being $\Delta$ the latent heat.

In the thermodynamic limit the energy distributions are
two delta functions situated at E$_a$ and E$_b$, then

\be
V_{\infty}^{min} = 1 - \frac{2(E_a^4 + E_b^4)}{3(E_a^2 + E_b^2)^2}
\label{EQ}
\ee

If we use V$_{\infty}^{min}$ = 0.66657 and E$_a$ in (\ref{EQ})
the value we got for the latent heat is $\Delta \approx 0.006$
which is of the same order as the one expected from the histograms.

\section{Conclusions}

The order of the Confinement-Higgs phase transition in the
SU(2)-Higgs model with fixed modulus is a highly non trivial issue.
We have used an extended parameter space, in order to get
a global vision on the problem. On this extended parameter
space we have found a line of first order phase transitions 
which get weaker
as $\kappa_2 \rightarrow$ 0. We have also observed that, because
of the computer resources needed, it is too ambitious trying
to measure two-peak energy distributions in the limit $\kappa_2$ = 0.
However, on this point, we can get conclusions 
from the behavior of the specific heat. 

As we have discussed along the paper, a fake second order PT
seems to exists for a range of $L$ in very weak first order
phase transitions.
We have applied Finite Size Scaling properties along this 
transient region to compute a $pseudo$ $\alpha/\nu$ critical index.
We want to be extremely careful at this point, this computation
is completely meaningless when the transition is first order, 
since Scaling does not hold, but it can be used
as a technical tool to catalogue the PT when there is no direct
evidences, as in this case.
Using the relation $\alpha = 2 - \nu d$, we got $\nu$ varying
in the interval (0.36, 0.41) in the range $L = 8, \dots, 20$. 
Calculated from $L$=20 and $L$=24, $\nu\approx0.35$.
We expect this behavior to be transitory, and when going to larger
lattices sizes, if the transition is second order,
$\nu$ should reach its mean field value $\nu=1/2$.
If the transition is first
order this value should go to 1/d, indicating that the specific 
heat maximum grows like the volume $L^d$. We believe that this is
the case, since the $pseudo$ $\nu$ exponent
in $L$=24, instead of approaching 1/2, starts to decrease.
An example of weak first order PT showing a similar
behavior is described in \cite{ISAF}.

In what concerning the motivation of introducing a second coupling,
we pointed out that $\kappa_2$ should not change 
the order of the PT because do not change the symmetry properties.
This argument is heuristic, but the phase
diagram we found supports this assertion.
As far as the order of the PT is concerned, we think that this approach
can be useful when dealing with PT of
questionable order in the sense that it is not clear whether
the transition is weakly first order or higher order. The hope
is that it could be applied to other more controversial models.

I thank A. Taranc\'on and L.A. Fern\'andez for comments and advice.

\end{document}